\documentclass[prl,twocolumn,showpacs,aps,floats,floatfix,superscriptaddress]{revtex4}

\usepackage[pdftex]{graphicx}
\usepackage{epsfig}

\begin{document}

\title{Prominence and control: The weighted rich-club effect}

\author{Tore Opsahl}\affiliation{School of Business and Management, Queen Mary College, University of London, UK}
\author{Vittoria Colizza}\affiliation{Complex Systems Lagrange Laboratory, Complex Networks, ISI Foundation, Turin, Italy}
\author{Pietro Panzarasa}\affiliation{School of Business and Management, Queen Mary College, University of London, UK}
\author{Jos\'e J. Ramasco}\affiliation{Complex Systems Lagrange Laboratory, Complex Networks, ISI Foundation, Turin, Italy}


\widetext

\begin{abstract}
Complex systems are often characterized by large-scale hierarchical organizations.  Whether the prominent elements, at the top of the hierarchy, share and control resources or avoid one another lies at the heart of a system's global organization and  functioning. Inspired by network perspectives, we propose a new general framework for studying the tendency of prominent elements to form clubs with exclusive control over the majority of a system's resources. We explore associations between prominence and control in the fields of transportation, scientific collaboration, and online communication. 
\end{abstract}


\maketitle 

Research has long documented the abundance of systems characterized by heterogeneous distribution of resources among their elements~\cite{Pareto:1897,Barabasi:1999}. Back in $1897$, Pareto noticed the social and economic disparity among people in different societies and countries~\cite{Pareto:1897}. This empirical regularity inspired the $80-20$ rule of thumb stating that only a select minority ($20\%$) of elements in many real-world settings are responsible for the vast majority ($80\%$) of the observed outcome. While recent studies have examined the tendency of prominent elements to establish connections among themselves~\cite{Colizza:2006}, how they leverage on their connections to gain and maintain control over resources circulating in a system still remains largely unexplored. In particular, do they collude and choose to exchange a disproportionately large amount of resources among themselves rather than with others? Or does competition prevent them from deepening the connections they have with one another? To answer these questions, we need to test for the tendency of prominent elements to engage in stronger or weaker interactions among themselves than expected by chance. We call this tendency the weighted rich-club effect. In this Letter, we adopt the framework of network theory~--~where the elements of the system are seen as nodes and the links among the elements represent interactions~\cite{Albert:2002,Doro:2003,Newman:2003,PastorVespibook,Caldarelli:2007}~--~ and provide a novel method to properly assess this tendency. 

Previous work has focused on highly connected nodes and the degree to which they preferentially interact among themselves~\cite{Colizza:2006}. This feature is known as the rich-club phenomenon~\cite{Colizza:2006,Zhou:2004}, a metaphor that alludes to the tendency of the highly connected nodes (i.e., the rich nodes) to establish more links among themselves than randomly expected. Evidence of the phenomenon has been reported for scientific collaboration networks~\cite{Colizza:2006}, transportation networks~\cite{Colizza:2006} and inter-bank networks~\cite{DeMasi:2006}. Conversely, research has shown that highly connected routers on the Internet tend not to be connected with one another~\cite{Colizza:2006}, whereas the pattern of interactions among proteins has been found to depend on the particular organism under consideration~\cite{Colizza:2006,Wuchty:2007}. Although uncovering interesting structural aspects of the systems, these studies are limited in that they only detect whether or not links among prominent nodes are present. In so doing, they neglect a crucial piece of information encoded in the weight of links, which is a measure of their intensity, capacity, duration, intimacy or exchange of services~\cite{Barrat:2004,Granovetter:1973}. A full understanding of how systems are organized  requires a shift towards a new paradigm that allows us to evaluate whether nodes that rise to network prominence also tend to exchange among themselves the majority of the resources flowing within the network. 

\begin{figure}
\begin{center}
\includegraphics[width=8.5cm]{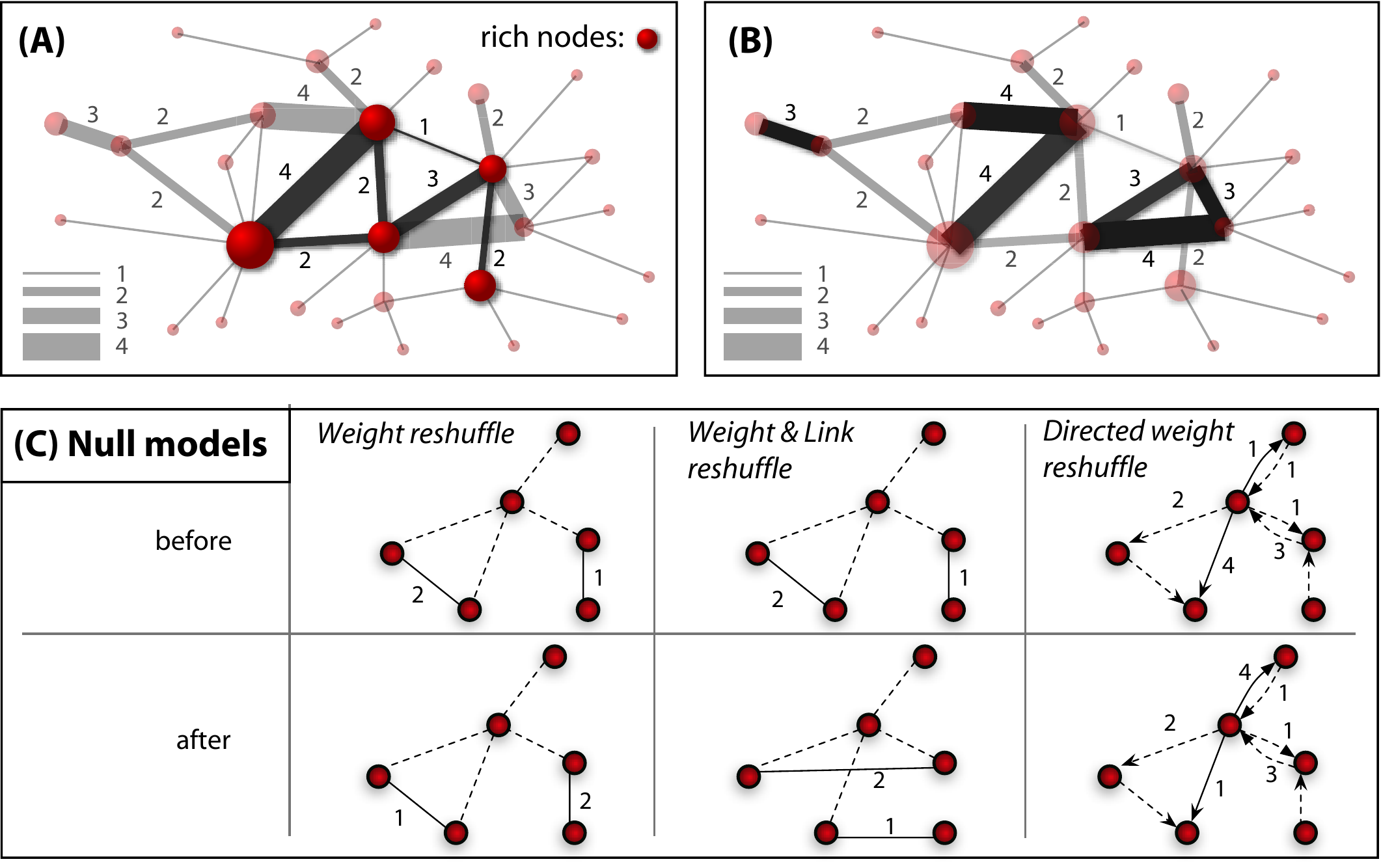}
\end{center}
\vspace{-0.5cm}
\caption{(A-B) Schematic representation of a weighted network, with size of
nodes proportional to their richness, and width of links to their weight
indicated by the corresponding numbers. Several definitions of richness can
be considered. (A) The nodes and links in the rich club are highlighted,
giving $E_{>r}=6$ links and $W_{>r}=4+2+2+3+1+2=14$. (B) The strongest
$E_{>r} = 6$ links of the network are highlighted, yielding the following
value for the denominator of Eq.~(\ref{eq:phiw}):
$\sum_{l=1}^{E_{>r}}w_l^{\mathrm{rank}}=4+4+4+3+3+3=21$. We thus obtain:
$\phi^w(r)=14/21$. (C) Null models. Solid lines refer to the links
reshuffled; the numbers  to their weight.}
\label{fig:wrc}
\end{figure}

To this end, we rank all nodes of a system in terms of a richness parameter
$r$. For each value of $r$, we select the group (the club) of all nodes
whose richness is larger than $r$. We thus obtain a series of increasingly
selective clubs. For each of these clubs, we count the number $E_{>r}$ of
links connecting the members, and measure the sum $W_{>r}$ of the weights
attached to these links (Fig.~\ref{fig:wrc}A). We then measure the ratio
$\phi^w(r)$ between $W_{>r}$ and the sum of the weights attached to the
$E_{>r}$ strongest links within the whole network (Fig.~\ref{fig:wrc}B).
Formally, we have:
\begin{equation}
\phi^w(r)=\frac{W_{>r}}{\sum_{l=1}^{E_{>r}}w_l^{\mathrm{rank}}}\,,
\label{eq:phiw}
\end{equation}
where $w_l^{\mathrm{rank}}\geq w_{l+1}^{\mathrm{rank}}$ with $l=1,\,2,\,...\,,\,E$ are the ranked weights on the links of the network, and $E$ is the total number of links. Eq.~(\ref{eq:phiw}) thus measures the fraction of weights shared by the rich nodes compared with the total amount they could share if they were connected through the strongest links of the network. Other measures can be introduced that depend on the local network structure surrounding the rich nodes~\cite{Colizza:2006,Serrano:2008,Valverde:2007,Zlatic:2008}. Here we aim at investigating the extent to which the prominent nodes control the flow of resources over the whole system.

In analogy with the topological rich-club measure~\cite{Colizza:2006,Amaral:2006}, a high value of $\phi^w(r)$, however, is not in itself sufficient to account for an actual tendency of the rich nodes to preside over the strongest links. This is due to the fact that even networks where links are randomly established could display a non-zero value of $\phi^w(r)$. To assess the actual presence of the weighted rich-club phenomenon, discounted of random expectations, $\phi^w(r)$ must be compared with an appropriate benchmark. To this end, we introduce a null model that is random but at the same time comparable to the real network. In particular, this model should break the associations between weights and links while preserving some crucial features of the network encoded in its degree distribution $P(k)$ (i.e., the probability that a given node is connected 
to $k$ neighbors) and weight distribution $P(w)$ (i.e., the probability that a given link has weight $w$). In addition, the nodes in the rich club must be the same as in the real network, which also preserves the richness distribution $P(r)$ (i.e., the probability that a given node has richness $r$).

In what follows, we introduce three procedures for constructing null models (see Fig.~\ref{fig:wrc}C) that correspond to different ways of preserving $P(r)$, depending on the choice of the richness parameter $r$. In this Letter, we explore three possible definitions of $r$: the degree $k$, the strength $s$ (i.e., the sum of the weights attached to the links originating from a node)~\cite{Barrat:2004} and the average weight $\bar{w}$ (i.e., the ratio between $s$ and $k$)~\cite{Ramasco:2006}. If the richness of a node is given by its degree, we adopt the following two randomization procedures. First, the Weight reshuffle procedure consists simply in reshuffling the weights globally in the network, while keeping the topology intact. Second, the Weight \& Link reshuffle procedure, which introduces a higher degree of randomization, consists in reshuffling also the topology, while preserving the original degree distribution $P(k)$~\cite{Newman:2003,Molloy:1995}. Weights are automatically redistributed by remaining attached to the reshuffled links. Both randomization procedures  can be easily generalized to directed networks. The Weight \& Link reshuffle procedure, mixing the signal coming from the topology with that generated by the location of weights, is considered here to assess the effects of higher degrees of randomization on the results, as well as for the sake of comparison with the topological rich-club coefficient~\cite{Colizza:2006}.

Inevitably, since weights are reshuffled globally, both procedures produce null models in which the nodes do not maintain the same strength $s$ as in the real network. When node richness is defined in terms of $s$, we need to introduce a third procedure that preserves this quantity. We construct a null model based on the randomization of directed networks~\cite{Serrano:2007} that preserves not only the topology and $P(w)$, but also the strength distribution $P(s)$ (i.e., the probability that a given node has strength $s$) of the real network. To this end, we reshuffle weights locally for each node across its outgoing links (Directed Weight reshuffle procedure). In so doing, we also obtain null models where the average weight $\bar{w}$ of outgoing links is kept invariant. We extend this procedure to the undirected case by duplicating an undirected link into two directed links, one in each direction. 

For a given definition of the richness $r$, the weighted rich-club effect can be detected by measuring the ratio:
\begin{equation}
\rho^w(r) = \frac{\phi^w(r)}{\phi_{\mathrm{null}}^w(r)}\,,
\label{eq:rho-global}
\end{equation}
where $\phi^w_{\mathrm{null}}(r)$ refers to the weighted rich-club effect assessed on the appropriate null model. When $\rho^w$ is larger than one, the original network has a positive weighted rich-club effect, with rich nodes concentrating most of their efforts towards other rich nodes compared with what happens in the random null model. Conversely, if it is smaller than one, the links among club members are weaker than randomly expected. 

In order to examine the applicability of our method, we study three real-world networks drawn from different domains: (i) The US Airport Network, obtained from the US Department of Transportation~\cite{BTS}, composed of $676$ commercial airports and $3,523$ routes connecting them. Each weight corresponds to the average number of seats per day available on the flights connecting two airports~\cite{Barrat:2004,Guimera:2005}. (ii) The Scientific Collaboration Network~\cite{Newman:2001}, extracted from the arXiv~\cite{arxiv} electronic database in the area of Condensed Matter Physics, from $1995$ to $1999$. Nodes represent scientists and a link exists between two scientists if they have co-authored at least one paper. Link weight reflects the authors contribution in their collaboration~\cite{Newman:2001}~--~the larger the number of authors collaborating on a paper, the weaker their interaction. (iii) The Online Social Network~\cite{Panzarasa:2007} comprising $59,835$ directed online messages exchanged among $1,899$ college students at the University of California, Irvine, from April to October $2004$. Link weight is the number of messages sent from one student to another.

\begin{figure}
\begin{center}
\includegraphics[width=8.5cm]{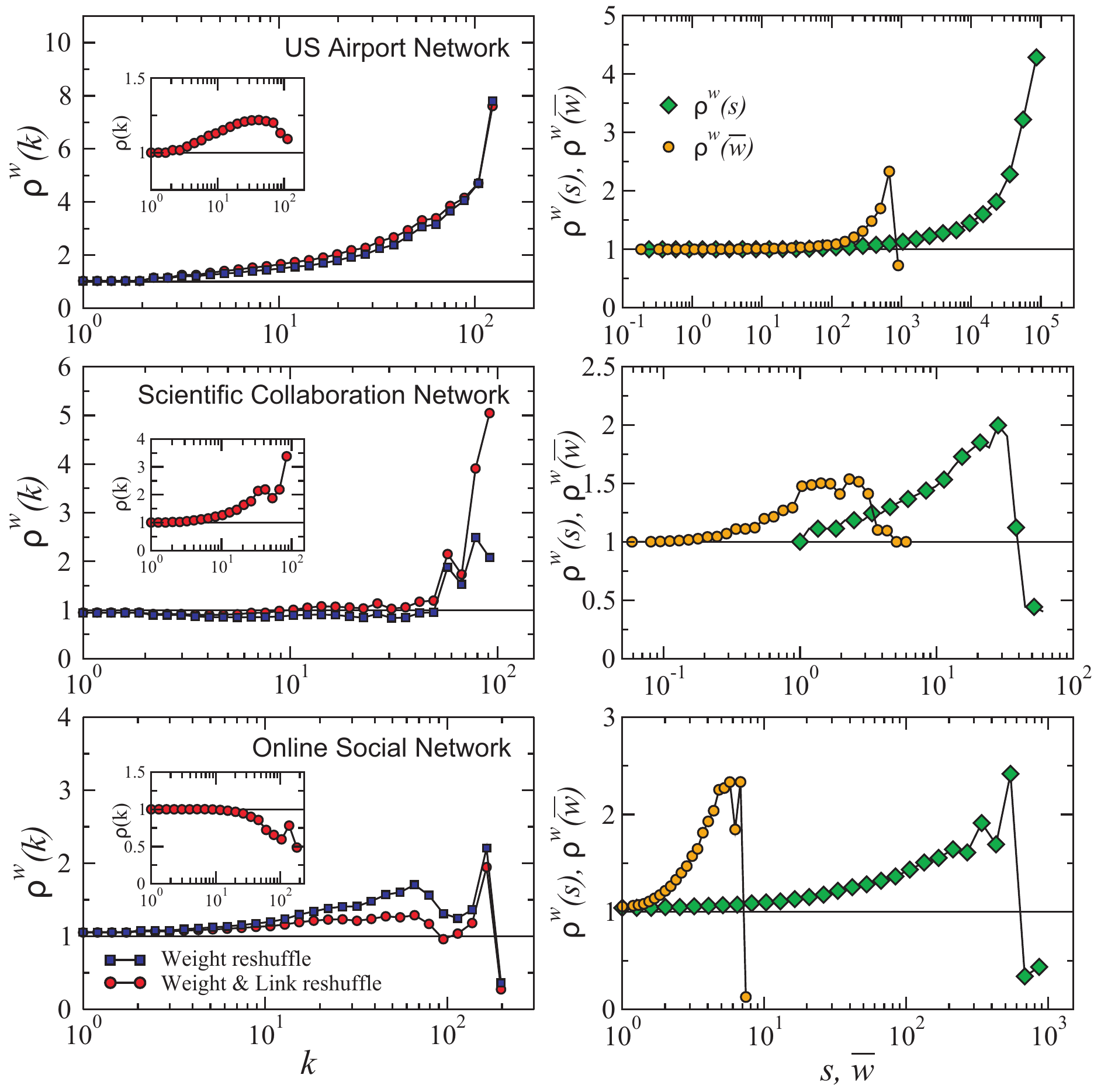}
\end{center}
\vspace{-0.5cm}
\caption{Weighted rich-club effect in: the US Airport Network (top); the Scientific Collaboration Network (center); and the Online Social Network (bottom). Left column: $r=k$. The insets refer to the topological rich-club coefficient $\rho(k)$~\cite{Colizza:2006}, defined as the ratio between $\phi(k)$ (i.e., the fraction of links connecting rich nodes, out of the maximum possible number of links among them)~\cite{Zhou:2004} and $\phi_{\mathrm{null}}(k)$ (i.e., $\phi(k)$ measured on the corresponding Weight \& Link reshuffle null model). Right column: $r=s$ (diamonds) and $r=\bar{w}$ (circles).}
\label{fig:wrc-ksw}
\end{figure}

We begin by defining network prominence in terms of node degree. In this case, $r=k$. We examine whether the highly connected nodes control the exchange of resources. For the three networks, Fig.~\ref{fig:wrc-ksw} (left column) reports the weighted rich-club ratio and its topological counterpart (inset). With only a mild topological effect~\cite{Colizza:2006}, the airport network shows a strong weighted rich-club effect, as can be identified from the remarkable growth of $\rho^w$ as a function of the degree of the airports. This finding agrees with previous studies that reported the presence of non-trivial correlations between weight of the links and degrees of the nodes~\cite{Barrat:2004,Guimera:2005,Wu:2006}. Connections among hub airports, with flights to many destinations, are characterized by large travel fluxes. Different results are found for the scientific collaboration network: while there is evidence of a strong positive topological rich-club effect, the network does not display a weighted one. As shown in Fig.~\ref{fig:wrc-ksw}, $\rho^w$ remains flat around $1$ for almost the whole range of $k$. The authors that have many collaborators tend to work together. However, the intensity of their collaboration does not differ from what is randomly expected, thus providing additional support to the observed lack of correlations between collaboration intensity and number of collaborators~\cite{Ramasco:2007a,Ramasco:2007b}. Finally, the weighted and topological rich-club effects display strikingly different trends for the online social network. Very gregarious individuals, with a large number of contacts, poorly communicate with one another. However, when they do, they choose to forge stronger links than randomly expected.

\begin{figure}
\begin{center}
\includegraphics[width=8.5cm]{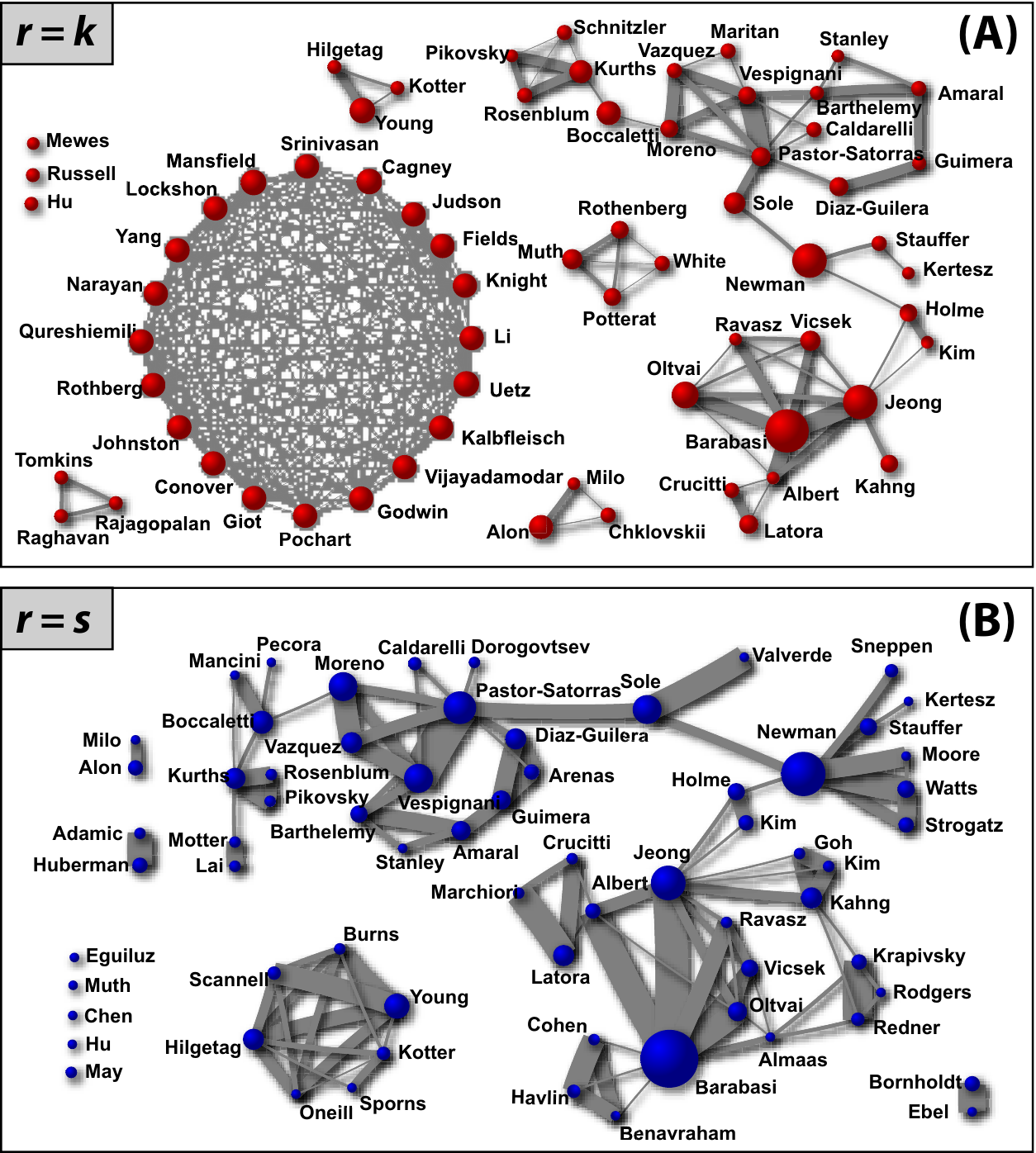}
\end{center}
\vspace{-0.5cm}
\caption{
Subsets of the rich nodes in the Network Science collaboration network~\cite{Newman:2006} based on degree (A) and strength (B). Only links among the rich nodes are shown. The size of the nodes is proportional to their richness; the width of the links  to their weight.
}
\label{fig:viz}
\end{figure}

To investigate how different definitions of prominence might affect the results, we restricted our attention to a subset of the arXiv collaboration network based on the publications on Network Science~\cite{Newman:2006}. In Fig.~\ref{fig:viz}A-B, we mapped the interaction patterns within the clubs obtained by defining $r$ in terms of the degree $k$ (number of co-authors) and the strength $s$ (number of published papers), respectively. In this network, each paper corresponds to a fully connected group of collaborators. When a paper is co-written by a large number of authors, these authors take on a high degree and thus increase their chances to become members of the club based on $k$. Large collaborations tend to secure club membership, yet generate weaker links than smaller ones~\cite{Newman:2001}. Experimental papers on biological networks are authored by a large number of scientists, and therefore only few such papers may suffice to substantially increase the topological rich-club effect (see the very large clique in Fig.~\ref{fig:viz}A). By contrast, they bring about weaker links than small-scale collaborations, thereby reducing their contribution to the weighted rich-club effect. However, when network prominence is defined in terms of $s$, club members as well as their interaction patterns substantially change with respect to the case in which $r=k$ (Fig.~\ref{fig:viz}B).

The next step is thus to define network prominence in terms of node strength $s$ and shift our attention from the most connected to the most involved nodes in the system's activity. Our findings show that active nodes preferentially direct their efforts towards one another, and this tendency becomes more pronounced as the involvement of nodes in network activity increases (see Fig.~\ref{fig:wrc-ksw}, right column). Not only do busy airports direct routes to one another, but they also secure control over travel fluxes by channeling on those routes a larger proportion of their passengers than randomly expected. This behavior is in sharp contrast with what was found using a different null model~\cite{Serrano:2008}, a pointed reminder of the crucial role played by such models in assessing the rich-club effect. When scientists are chosen on the basis of their scientific productivity $s$, exclusive clubs emerge in which scientists tend to collaborate with one another to a greater extent than randomly expected, unlike what was found within the club of the most connected scientists. In the online social network, highly active users tend to communicate with one another more frequently than would be the case if contacts were chosen at random. 

While node strength gives a general indication of how involved a node is in the activity of a network, it does not allow us to discriminate between nodes with a large number of weak links and nodes with a small number of strong links. To address this issue, we define the richness parameter in terms of the average weight $\bar{w}$. We find positive effects in all networks (see Fig.~\ref{fig:wrc-ksw}, right column). Airports that optimize the traffic per link tend to direct their busy routes to one another. Scientists that show the ability to maximize their resources per collaboration  tend to intensively collaborate with one another. Online communication tends to occur among users that maximize the attention directed to each contacts.

By shifting focus from the network topology to the weight of links, we have proposed a new general framework for studying the tendency of prominent nodes to attract and exchange among themselves the majority of the resources available in a system. Unlike a merely topological assessment of the network, our method allows us to uncover organizing principles that would otherwise remain undetected. In addition, by varying the definition of prominence, we found evidence of different organizing principles, and paved the way towards a deeper understanding of the multiple layers of a system's organization. Our method is widely applicable, in that it enables us to study the control benefits of prominent elements in a variety of empirical settings, by decoupling prominence from strictly network properties. To the extent that the components of a system can be sorted into a hierarchy according to a given property, our framework suggests several new ideas for future research, including the impact of performance, centrality, status, age, size on the ability of elements to control flows of resources. In this respect, our study helps shed a new light on the global organization of complex systems.

\begin{acknowledgments}
The authors would like to thank Alessandro Vespignani and Alain Barrat for useful discussions and suggestions.  
\end{acknowledgments}


\end{document}